# Modeling and Simulating Agent-Based City Migration Using Conway's Game of Life

Bruce Deng, Mayank Kejriwal[1]

[1]Corresponding author: kejriwal@isi.edu

# Abstract

Agent-based modeling (ABM) has become a cornerstone of complexity science, enabling the study of heterogeneous agents interacting within dynamic environments. Among ABM frameworks, John Conway's Game of Life (GoL) stands out for its simplicity and ability to generate emergent macroscopic patterns from basic microscopic rules. In this paper, we propose and implement a novel GoL-based framework to simulate urban migration dynamics. Using a grid-within-a-grid approach, our approach encodes probabilistic tendencies for out-migration due to densification and sparsification, simulating the evolution of population centers. By initializing GoL grids with different distributions and parameterizing migration preferences, we explore how urban structures emerge and stabilize over time. Through a series of experiments, we demonstrate that even with simple rules, this framework shows promise for understanding emergent urban phenomena, providing insights into city growth and structure. Methodologically, our framework offers a versatile and computationally efficient tool for studying urban migration patterns, contributing to the broader application of ABMs in computational urban social science.

# Introduction

Starting from the early 2000s, *agent-based modeling* (ABM) has emerged as an important paradigm in complexity science (Bankes, 2002; Helbing, 2012). ABM is a powerful computational approach for studying complex systems characterized by heterogeneous agents interacting within dynamic environments (Gilbert, 2019). By simulating individual behaviors and localized decision-making processes within such a framework, researchers can better investigate emergent phenomena that are difficult to capture (or even formulate) using traditional and closed-form analytical methods (Macy and Willer, 2002). AGM has been used extensively in computational social science to investigate phenomena ranging from ecological dynamics (Janssen and Ostrom, 2006) and social networks to modeling economic markets and healthcare systems (Arthur, 2006; Hamill and Gilbert, 2015; Cassidy et al., 2019).

While ABMs can themselves become complicated and large-scale in design, there continues to be interest in frameworks that are governed by relatively simple and established *microscopic* rules. John Conway's *Game of Life* (GoL) platform is one such example (Conway, 1970). We describe the rules of GoL subsequently, but note here that it has been applied in fields ranging from understanding the development of complexity in early life and other biological processes, to agent-based simulations of economic phenomena (Caballero, Hodge, and Hernandez, 2016; Del Faro-Odi, 2020; Faux and Basson, 2023). Its core strength is that it can be used to create interesting self-replicating patterns at a *macroscopic* level despite the basic microscopic rules governing each agent's local behavior.

One such complex phenomenon is the evolution of population centers like cities using the kinds of simple and 'local' rules that serve as a good probabilistic representation of migration-related and economic phenomena like out-migration due to extreme densification (e.g., a tendency of people to migrate out of densely populated regions to suburbia) and sparsification (e.g., a tendency of people to migrate out of more sparsely populated regions due to, for example, lack of opportunities or resources). We say *tendency* because our goal is to model and understand the macroscopic effects of such preferences (expressed probabilistically) at a larger scale using the simple rules of GoL. In the literature, there is precedent for such tendencies through a slew of findings. For example, Del Faro-Odi et al. (2020) show how GoL can be used to reproduce the statistical properties of wealth distributions (which tend to obey power law-like distributions) observed in real economic data. In this context, Schmickl (2022) discusses the concept of strong emergence arising from weak weak emergence.

By encoding tendencies or preferences, and simulating GoL starting from an initial distribution, we can observe how the system evolves (especially under different parameterizations of both the initial board and the encoded preferences) over time. With this motivation in mind, we implement and simulate a simple, yet scale-invariant, *agent-based city migration* framework, based on the rules of GoL. We use the framework to quantify macroscopic and emergent properties that are consistently observed after the simulation has had time to converge, or otherwise been executed for sufficiently long time horizons.

Empirically, through a series of brief experiments, we show that even with simplistic rules and a reasonable ring-based model where GoL grids are 'nested' within an outer grid, agent-based simulations can effectively capture and model the complex interplay of individual decisions and environmental constraints on a city's growth and structure. Methodologically, our framework offers a scalable and adaptable approach for simulating complex, emergent behaviors in urban migration patterns using an ABM.

# Experimental Setup

## Conway's Game of Life (GoL)

*Game of Life* (GoL) is a cellular automaton created by British mathematician John Conway over 50 years ago (1970). The game simulates a two-dimensional grid of cells, each cell being alive or dead. Each cell has eight neighbors; in each timestep of the simulation, the status of every cell's neighbors will be evaluated. The rules are as follows:

- if an alive cell has exactly two or three, neighbors, it will stay alive;
- if an alive cell has less than two or greater than three neighbors, it will die by solitude or overpopulation, respectively;
- if a dead cell has exactly three neighbors, it will become alive.

As alluded to earlier in *Introduction*, many studies in complexity science, and research areas intersecting with it, have used GoL (or a close variant) as a simulation platform for their investigation. In this paper, we use the usual model without any changes. Given a specific-sized board, the key variable that we attempt to control in the model is the initialization of the grid i.e.,a determination of which cells are alive, and which cells are alive, at time *t=0*.

## Pilot Experiment for Validating GoL Implementation

To efficiently run controlled experiments, we used standard Python packages like NumPy and Matplotlib to implement GoL. We used two NumPy arrays to represent the grid. As the array containing the grid is being iterated through, any changes made (a cell becoming alive or dead) are made to a separate, identical array to avoid modifications affecting the results of other cells during the process. Then, the original grid is set to the grid where the changes were made.

We validated our implementation through a 'pilot' experiment that uses a 10x10 grid. We sought to measure the impact that the probability of spawning an alive or dead agent in any given cell in the initialization of the grid would have on the convergence of the simulation (assuming that it converged). A simulation is said to converge at time *t* in this context when the grid configuration in time *t* of the simulation is the same as the grid configuration in time[1] *t+1*. For a small grid like a 10x10, convergence is usually measurable, but at larger scales, convergence is almost always unattainable. We limited the number of timesteps for each trial of the experiment to 1000. To determine whether a cell in the grid starts off as alive or dead, we took as user-input a probability, which was used to randomly determine the initial status of the cell. If the probability is higher, it is likely that a large proportion of the cells will begin as alive. We ran fifteen trials of the experiment for each probability value, ranging from 0.1 to 1.

---

[1] It is straightforward to verify, given the rules of Game of Life, that if this condition is met, it is necessarily the case that the configuration in time *t+2* will be the same as that in time *t+1* (and *t*) and so on.

Figure 1 (top) displays the number of timesteps needed on average for convergence to be achieved on different probability values, ranging from 0.1 to 1. Each point on the graph represents the average value of approximately 15 trials. For trials where convergence was not achieved, the data point was excluded from the calculation to avoid outliers affecting the calculation of the mean. Evidently, we see that at extremely small or large probabilities of a cell being alive, the timesteps required to converge are much shorter than probabilities that vary around 0.5.

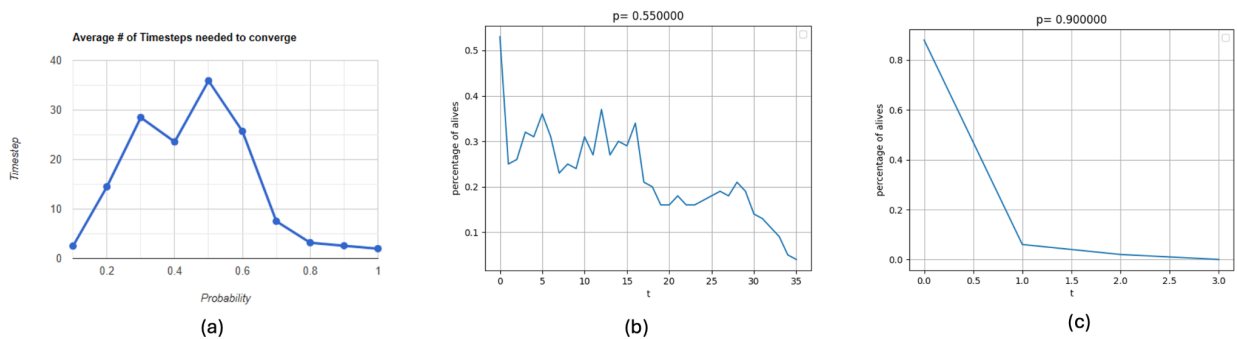

**Figure 1:** (a) the number of timesteps needed on average for convergence to be achieved on different probability values, ranging from 0.1 to 1. Each point on the graph represents the average value of approximately 15 trials. For trials where convergence was not achieved, the data point was excluded from the calculation to avoid outliers affecting the calculation of the mean. (b) and (c) display the percentage of cells that are alive throughout the course of the simulation for different probability values.

Figures 1b and c display the percentage of cells that are alive throughout the course of the simulation for two different probability values. In both cases, convergence was achieved within the timesteps shown on the graph. An interesting empirical finding was that, for large values of probability, the percentage of cells that were alive quickly depleted within one or two timesteps due to overpopulation. Meanwhile, for a middle-sized probability value, that process took much longer, and the percentage of alive cells would fluctuate up and down at times.

As further validation, we extended the size of the grid from 10 x 10 to 100 x 100. The methodology for this experiment generally stayed the same as the previous one, except using arrays that held 100x more cells. We again capped the timesteps to 1000; if the simulation did not converge by then, the program would be stopped.

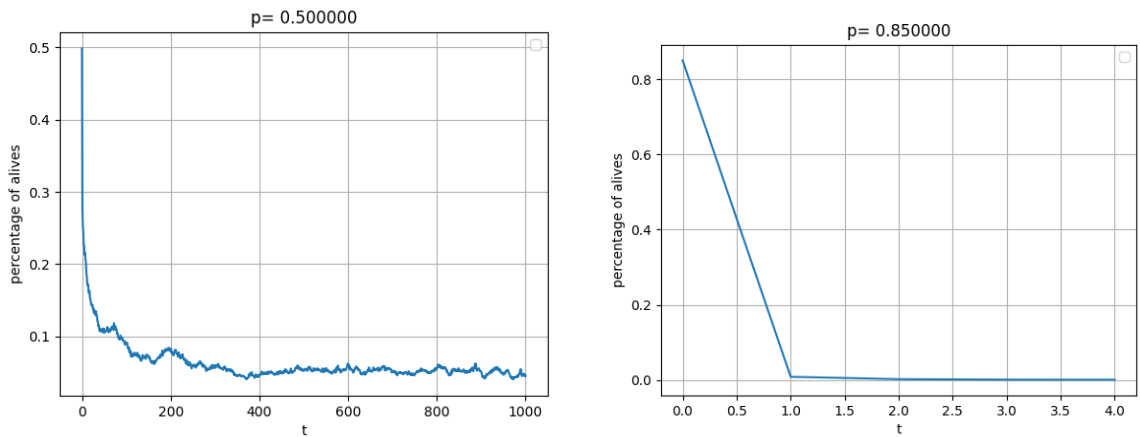

**Figure 2:** The percentage of alive cells on the 100x100 board when the probability value was set to 0.5 (left) and 0.85 (right).

Figure 2 (left) depicts the percentage of alive cells on the 100x100 board when the probability value was set to 0.5. Unlike the results depicted in Figure 1, there was a steep decline in alive cells without any fluctuation up until the percentage of alive cells reached 0.1. Then, the percentage of alive cells fluctuated around the 0.05 mark until the completion of the simulation, without ever reaching convergence. The inability for the simulation to reach convergence was generally expected at these probability levels due to the large number of cells on the grid.

In Figure 2 (right), the graph is similar. We see a sharp decline in the percentage of alive cells within one timestep, and by the fourth timestep, the simulation is able to converge. Therefore, it can be concluded that at high values of the probability of a cell being alive upon initialization, the simulation will converge, regardless of the size of the grid.

## Ring Agent-Based City Migration Model

For investigating migration-based phenomena in a grid of *variable* density, we propose an agent-based city migration model that uses a representation of *rings* or concentric circles to model areas of variable density within a grid ('city'). Essentially, the grid is divided into distinct rings of decreasing population density as one moves further away from the center of the grid. This model is able to simulate and capture patterns of simplified urban population density gradients, with the innermost concentric circle rings containing higher occupancy, or alive cell-rates than the sparser outer concentric rings.

More formally, we consider a *grid within a grid* model where we first assume an *outer* grid of size N X N, where *N* is indivisible by 2. Each cell in this grid is then instantiated as a grid of size N-1 X N-1. Concentric rings are determined by their distance from the center cell, located at $\left(\frac{N-1}{2}, \frac{N-1}{2}\right)$. We associate each ring *k* with a (generally) different probability of initial cell aliveness ($P_k$), which is simply the probability of a cell being alive in each ring *k*. We set the sequence of probabilities to decrease gradually, with the central ring (comprising just one 'inner' grid) being the most dense. For example, if there are five rings, we would set $P_1$ = 0.9, $P_2$ = 0.8, $P_3$ = 0.6, $P_4$ = 0.4, $P_5$ = 0.2, where $P_1$ corresponds to the ring closest to the center and $P_5$ corresponds to the ring farthest from the center.

## Results and Discussion

Figure 3 shows the results of the simulation by plotting the percentage of alive cells versus time in the 110 X 110 inner grid model. The figure shows that the higher probability values (orange, green, and red plots) are the most volatile, as they constantly fluctuate between sudden increases and decreases, while the changes for mid-probability values (brown and purple plots) are not as dramatic. As expected, the lowest probability value of 0.2 (blue plot) exhibits behavior similar to that of the mid-probability trends, since probability values of 0.4 and 0.2 are the two most frequent in the entire grid. Additionally, for the highest probability value, the plot ends at a smaller percentage of alive cells than the other four lines, which may be due to the fact that many of the cells will die by overpopulation, as approximately ninety-percent of the cells in the orange ring will be initialized as alive. Figure 4 displays the overall proportion of cells that are alive among the entire grid and suggests a similar finding even when averaging across probability values.

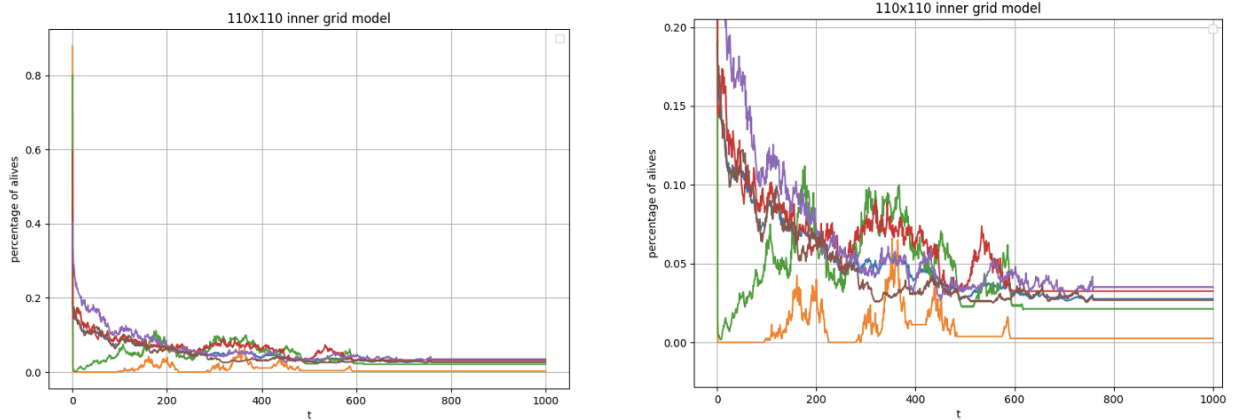

**Figure 3:** Percentage of alive cells versus time in the 110 X 110 inner grid model. The right figure is a y-axis magnified view of the left figure after accounting for the high outliers on the y-axis for early values of t. In both figures, the orange, green, red, purple, and brown plots indicate probability values of 0.9, 0.8, 0.6, 0.4 and 0.2, respectively.

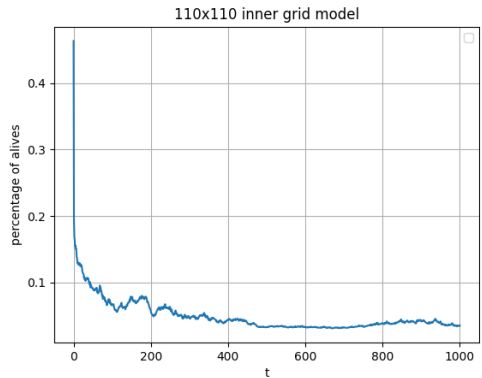

**Figure 4:** The overall proportion of cells that are alive in the entire grid versus time.

Figure 5a depicts a visualization of the *Agent-based City Migration Model* in the last timestep ($t$ = 1000). At this point, the "gradient" of decreasing alive cells in the concentric rings that appear farther from the center is quite nonexistent; the proportion of alive cells in every concentric ring is around the same ($P \approx 0.05$). We also observe a variety of patterns, including crosshairs, squares, and gliders, being created by the interactions between the cells. The bottom two figures depict a similar visualization, but for grids with uniform probability of a cell being alive (*P* = 0.5) of size 10 x 10 and 100 x 100.

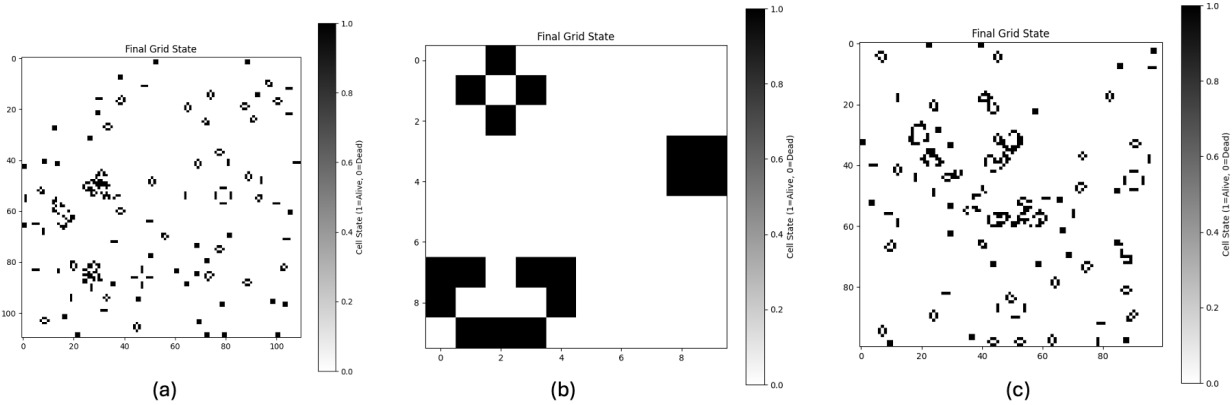

**Figure 5:** (a) depicts a visualization of the *Agent-based City Migration Model* in the last timestep (*t* = 1000). (b) and (c) depict similar visualizations, but for grids with uniform probability of a cell being alive (*P* = 0.5) of size 10 x 10 and 100 x 100, respectively.

Together, these results show that, even though the inner grid is the most dense at the beginning of the simulation, on average, the entire grid starts to sparsify even after short time horizons. The effects of overcrowding start to take hold in regions close to the center. The ring furthest from the center continues to stay sparse. Higher density regions (although still relatively sparse) are instead observed between the 'inner city' and the 'suburbs' where the death and alive rates eventually balance out. Interestingly, pockets of high density are still observable. Although random, such pockets always emerge somewhere in the grid. This phenomenon provides a simple representation of the types of complex congregation and gentrification patterns we sometimes see in cities. Here, such effects are arising because of a *controlled snowballing* phenomenon (a few closely clustered alive cells lead to more alive cells, but if there are too many alive cells, the death rate starts to climb again); in real cities, the effects are likely due to more exogenous causes (for example, when a large company decides to build its next headquarters or an office in a neighborhood).

## Conclusion and Future Work

This study demonstrates the utility of a Game of Life (GoL)-based framework for modeling urban migration dynamics through simple, probabilistic rules. By encoding tendencies such as out-migration due to densification or sparsification using a nested, ring-based grid structure, we suggested how complex, emergent behaviors in urban population distributions can be simulated and analyzed. The scalability and adaptability of the approach allowed us to explore various spatial configurations and parameterizations, offering insights into the macro-level effects of micro-level urban agent decisions. Our findings reinforce the potential of ABMs with foundational rule sets to provide meaningful insights into real-world phenomena, such as urban growth and structural patterns in city migration dynamics.

Future work will focus on extending the framework in several directions. First, incorporating more detailed agent behaviors, such as economic, social, and environmental factors influencing migration decisions, could improve the model's fidelity. Second, the introduction of stochastic external shocks, such as natural disasters or policy changes, could further test the robustness of emergent patterns. Third, integrating real-world data to validate and refine the model would bridge the gap between simulation and observed urban migration phenomena. Finally, developing visualization and optimization tools for decision-makers could improve the framework's practical applicability in urban planning and policy-making.